\documentclass[showpacs,aps,floatfix,prl,reprint,superscriptaddress]{revtex4-1}
\usepackage{xfrac}
\usepackage{amssymb}
\usepackage{braket}
\usepackage{graphicx}
\usepackage{verbatim}
\usepackage{etoolbox}
\usepackage{amsfonts} 
\usepackage{amsmath} 
\usepackage[utf8]{inputenc} 
\usepackage{mathtools}
\usepackage{soul,xcolor,colortbl}
\usepackage{bm} 
\usepackage{array}
\newcolumntype{C}[1]{>{\centering\arraybackslash}p{#1}}\usepackage{soul}
\definecolor{Gray}{gray}{0.85}

\usepackage{color}  
\def\ACBN{{ACBN0}}  
\begin{document}

\setstcolor{red}

\title{Reformulation of DFT+$U$ as a pseudo-hybrid Hubbard density functional for accelerated materials discovery}
\author{Luis A. Agapito} 
\affiliation{Department of Physics and Department of Chemistry, University of North Texas, Denton, TX 76203, USA}
\affiliation{Center for Materials Genomics, Duke University, Durham, NC 27708, USA}
\author{Stefano Curtarolo}   
\affiliation{Center for Materials Genomics, Duke University, Durham, NC 27708, USA}
\affiliation{Materials Science, Electrical Engineering, Physics and Chemistry, Duke University, Durham, NC 27708, USA}
\author{Marco \surname{Buongiorno Nardelli}} 
\email{mbn@unt.edu}
\affiliation{Department of Physics, University of North Texas, Denton, TX 76203, USA}
\affiliation{Center for Materials Genomics, Duke University, Durham, NC 27708, USA}
\date{\today}

\begin{abstract} 
  The accurate prediction of the electronic properties of materials at a low computational expense is a necessary conditions for the development of effective high-throughput quantum-mechanics (HTQM) frameworks for accelerated materials discovery.
  HTQM infrastructures rely on the predictive capability of Density Functional Theory (DFT), the method of choice for the first principles study of materials properties. However, DFT suffers of approximations that result in a somewhat inaccurate description of the electronic band structure of semiconductors and insulators. 
In this article we introduce \ACBN, a pseudo-hybrid Hubbard density functional that yields an improved prediction of the band structure of insulators such as transition-metal oxides, as shown for TiO$_2$, MnO, NiO and ZnO, with only a {\it negligible} increase in computational cost.

\end{abstract}
\maketitle 
\section{introduction} 

\emph{High-Throughput Quantum-Mechanics} (HTQM) computation of materials properties by \emph{ab initio} methods has become the foundation of an effective approach to materials design,  discovery and characterization \cite{nmatHT}. 
This data-driven approach to materials science currently presents the most promising path to the development of advanced technological materials that could solve or mitigate important social and economic challenges 
of the 21st century \cite{MGI_OSTP,nmatHT_editorial,Ceder_ScientificAmerican_2013,Greeley2006,curtarolo:TIs,Fornari_Piezoelectrics_PRB2011,aflowSCINT,aflowTHERMO,YuZunger2012_PRL,curtarolo:art86,nmatHT}.
In order for this approach to be successful, however, one needs the confluence of three key factors: 
{\bf i.} improved computational methods and tools;
{\bf ii.} greater computational power; and
{\bf iii.} heightened awareness of the power of extensive databases in science\cite{nmatHT_editorial}.
While the last two are driven by technological advances in computing and research and development needs, the development of improved computational tools appropriate for HTQM frameworks is a grand challenge that is still in need of considerable advances. 

Most HTQM infrastructures rely on the predictive capability of Density Functional Theory (DFT), the method of choice for the first principles study of materials properties. However, despite the enormous success of DFT in describing many physical properties of real systems, its limitations in describing correctly the electronic band structure of insulators are well known. The method is limited by the presence of an unknown correlation term that represents the difference between the true energy of the many-body system of the electrons and the approximate energy that we can compute. The common approximations based on a nearly-homogeneous-electron-gas treatment of the electron density, the local-density (LDA) and generalized-gradient approximation (GGA), are extremely successful in the description of many  physical properties of materials, but dramatically underestimate the electron energy gap in insulators and semiconductors and thus fail to satisfactorily describe the electronic properties of these systems. Higher order levels of theory that are able to predict with great accuracy the energy gaps exist (the $GW$ approximation\cite{Hedin_GW_1965} and dynamical mean-field theory\cite{Georges1992Hubbard_model_infinite,Georges1996DMFT,Kotliar2006Electronic_DMFT}, among others) but they are computationally expensive and unsuitable for extensive high-throughput materials characterization \cite{monsterPGM,curtarolo:art96},
even when Machine Learning methods are employed to simplify the complexity of the task especially when seeking for new materials systems \cite{curtarolo:art84,curtarolo:art85}.
In order to address the energy gap problem at a lower computational cost the two most common corrections to traditional local and non-local approximations to DFT are  ``hybrid functionals'' and DFT+$U$. 
Both approaches aim to reduce at some level the self-interaction error \cite{PerdewZunger} and introduce the derivative discontinuity in the exchange-correlation functional\cite{Perdew83, Sham83, Perdew1982DFT_for_fractional} responsible for the underestimation of the energy gap\cite{Ivady2014Theoretical_unification}.

Hybrid functionals are based on the idea of computing the exact exchange energy from the Kohn-Sham wavefunctions and to mix it with the (semi)local approximation of exchange energy of DFT\cite{Kummel2008OrbitalDFT}. The method is very successful in predicting the energy gap with good accuracy and points to the importance of introducing some degree of exact exchange in traditional DFT for a proper description of the electronic structure. However, the level of mixing is not determined from first principles, so the method suffers of some level of empiricism and, even after the introduction of range separated functionals suitable for periodic system calculations\cite{Heyd2003,HSE_2006_JCP}, it is more computationally demanding than LDA or GGA. 

The treatment of systems with strongly localized(correlated) electrons is another  outstanding issue that limits the predictive power of (semi)local approximations to DFT for systems where localization is important, such as transition metal insulators, and directly impacts the energy gap problem. The DFT+$U$ method introduced by Liechtenstein and Anisimov \cite{Liechtenstein1995,Anisimov1997} aims at preserving the information of orbital localization from being averaged out as in LDA or GGA in order to improve the description of the band structure with a very modest increase in computational effort. Since in this paper we are mainly concerned with an extension of this approach, let us discuss its guiding principles in some detail.


Within the DFT+$U$ ansatz, localized states $\varphi_i$ largely retain their atomic nature and, therefore, can be expanded in term of an atomic-orbital basis set $\{\phi_m \} \equiv \{m\}$. The Coulomb and exchange energy associated with these states 
is explicitly evaluated using the Hartree-Fock (HF) framework via electron repulsion integrals (ERI,
also know as two-electron integrals) with a screened (renormalized) Coulomb interaction $V_{ee}$ and added to the original DFT energy after the removal of double counting terms in the energy expansion, in a spirit similar to the hybrid functionals approach. 

The HF Coulomb and exchange energy of the localized states is given by \cite{Liechtenstein1995}:
\begin{align} \label{eq:liechtenstein}
E^{\{m\}}_{\textrm{HF}}&=\frac{1}{2} \sum_{\{m\},\sigma}
\{ \langle mm'' | V_{ee} | m'm''' \rangle n^{\sigma}_{mm'} n^{-\sigma}_{m''m'''} \notag \\
&+ \left( \langle mm'' | V_{ee} | m'm''' \rangle - \langle mm'' | V_{ee} | m'''m' \rangle \right) \notag \\
&\times n^{\sigma}_{mm'}n^{\sigma}_{m''m'''} \},
\end{align}
where $n^{\sigma}$ is the spin density matrix $n^{\sigma}$ of the atomic orbitals $\phi_m$.
This equation can be simplified via the introduction of the  phenomenological parameters $\bar{U}$ and $\bar{J}$ that describe the on-site Hubbard-like interactions as expressed by Dudarev \textit{et al.} \cite{Dudarev1998}:
\begin{align} \label{eq:dudarev}
E^{\{m\}}_{\textrm{HF}}&\approx\frac{\bar{U}}{2} \sum_{\{m\},\sigma}
N^{\sigma}_{m} N^{-\sigma}_{m'} + \frac{\bar{U}-\bar{J}}{2} \sum_{m \neq m',\sigma} N^{\sigma}_{m} N^{\sigma}_{m'}.
\end{align} 
Here, $N^{\sigma}_{m}$ is the spin occupation number of the atomic orbital $\phi_m$.

From the equations above, it clearly follows 
that the new parameters, $\bar{U}$ and $\bar{J}$, contain the information of all the ERIs in an averaged scenario.
In physical terms, $\bar{U}$ is the strong correlation experienced between localized electrons {---} only subtly coupled to the sea of extended states {in which} they live. Thus, the most akin definition of $\bar{U}$ (for the non-spin-polarized case) is the average \cite{Shih_screened_coulomb_interaction_PRB2012}:
\begin{equation} \label{eq:d2}
\bar{U}=\frac{1}{(2l+1)^2}\sum\limits_{i,j} \langle \varphi_i \varphi_j | V_{ee} | \varphi_i \varphi_j \rangle,
\end{equation}
 {where} $2l+1$ {is} the total number of localized states $\varphi_i$, {and} $l=2,3$ for $d,f$ orbitals, respectively. 
 {The exchange contribution, $\bar{J}$, is given by a similar average \cite{Shih_screened_coulomb_interaction_PRB2012}.}

Although the physical picture is clear, an unambiguous procedure for computing $\{\bar{U},\bar{J}\}$ from {\it ab-initio} {does not exist}.
Two factors need to be further clarified in Eq.~\ref{eq:d2}: \textit{i}) the screened (renormalized) Coulomb interaction $V_{ee}$ arising from the ``subtle coupling'' to the background extended states; and \textit{ii}) the actual orbitals $\varphi_i$ used to represent the ``localized electrons''.  

Amongst the most common {\it ab-initio} methods to compute $\bar{U}$ are the
constrained random-phase approximation (cRPA) \cite{Springer_frequency_interaction_NI_RPA_PRB1998} and the linear-response constrained DFT (or cLDA) \cite{Gunnarsson_dft_anderson_Mn_CdTe_PRB1989,Cococcioni_DFTU_LinearResponse_PRB2005}.
 {The former} computes the screened Coulomb interaction as the bare Coulomb interaction renormalized by the inverse dielectric function, which is calculated using the random phase approximation.
 {The latter} circumvents the ambiguity of $V_{ee}$ by indirectly determining $\bar{U}$ as the second derivative of the total energy with respect to constrained variations of the atomic charge $q_I$ of the chosen Hubbard center $I$, $\bar{U}={\partial^2 E}/{\partial {q_I}^2}$.  
$E$ is the total energy of a supercell large enough to converge to the bulk environment for the atom $I$. It is assumed that the charge perturbation on atom $I$ does not disturb the local environment. In DFT with linear-combination-of-atomic-orbital (LCAO) basis, this is enforced by suppressing the hopping integrals to prevent charge rehybridization or transfer with its environment \cite{Gunnarsson_dft_anderson_Mn_CdTe_PRB1989} and in the case of plane-wave DFT by subtracting a correcting term from ${\partial^2 E}/{\partial {q_I}^2}$    
 {as} given in Ref.~\onlinecite{Cococcioni_DFTU_LinearResponse_PRB2005}. 
This method has been widely used for open-shell systems; nonetheless, the numerical reliability becomes {challenging} 
for closed-shell systems where the localized bands are completely full, {thus exhibiting} very small response to the linear perturbation \cite{Lee_linear_response_Cu_Zn_ZnO_KOREA2012,Cococcioni_reviewLDAU_2014}.

Regarding the representation of the $\varphi_{i}$ states, {\textit e.g.} $d$ or $f$ electrons of transition metals, localized orbitals 
---obtained either from the linear-muffin-tin-orbital (LMTO) method \cite{Aryasetiawan_muffin_tin_PRB1994} or from the $N^{th}$-order muffin-tin-orbital ($N$MTO) method \cite{Andersen_muffintin_PRB2000}---
can be used with both the cLDA and cRPA to obtain $\bar{U}$.
Recently, maximally localized Wannier functions (MLWF), {an invariant choice} suitable for plane-wave calculations, have also been employed \cite{Miyake_effective_models_PRB2009,Shih_screened_coulomb_interaction_PRB2012}. 
By construction, these {functions} 
are associated with a given angular momentum ($l,m$) {and the direct correspondence makes them convenient} to represent the localized $d$ or $f$ electrons. 
 {Ultimately,} pinpointing a single localized state within the solid is arbitrary --- any of these options are equally valid.
In principle, {the options should be equivalent} for very localized states; nonetheless, the physical significance and construction becomes more ambiguous when bands corresponding to localized states are not fully disentangled \cite{Miyake_effective_models_PRB2009}. Despite attempting the computation of the same physical entity, the cLDA and cRPA methods 
do not yield the same value of $\bar{U}$ \cite{Aryasetiawan_DFTU_PRB2006}. Considering the number of assumptions taken in numerical implementations, {the outcome is unsurprising.}

In this article, we introduce an alternative {\it ab-initio} method to compute $\bar{U}$ and $\bar{J}$, which parallels the calculation of the HF
energy for molecules and solids and follows closely the original definition of Anisimov {\it et al.} (Eq.~\ref{eq:liechtenstein}):
 the \underline{A}gapito-\underline{C}urtarolo-\underline{B}uongiorno \underline{N}ardelli (\ACBN) pseudo-hybrid Hubbard 
density functional.
 In \ACBN\, the Hubbard energy of DFT+$U$ is calculated via the direct evaluation of the local Coulomb ($\bar{U}$) and exchange ($\bar{J}$) 
integrals in which the screening of the bare Coulomb interaction is replaced by a renormalization of the density matrix.
 Through this procedure, the values of $\bar{U}$ and $\bar{J}$ are thus functionals of the electron density and depend directly on the chemical environment and crystalline field, introducing
 an effective procedure of giving the proper description of Mott insulators and other strongly correlated transition-metal oxides. 
As a first application, we discuss the  electronic properties of a series of transition metal oxides that show good agreement  with hybrid functionals, the
 $GW$ approximation and experimental results at a fraction of the computational cost. In particular, we will demonstrate that the \ACBN\ functional 
satisfies the rather ambitious criteria outlined by Pickett {\it et al.} in one of the first seminal {articles} on LDA$+U$ \cite{Pickett1998}: 
\textit{i}) \ACBN\, reduces to (LDA)PBE when (LDA)PBE is known to be good; 
\textit{ii}) the energy is given as a functional of the density; 
\textit{iii}) the method specifies how to obtain the local orbital in question; 
\textit{iv}) the definition of $\bar{U}$ and $\bar{J}$ is provided unambiguously.
and \textit{v}) the method predicts antiferromagnetic insulators when appropriate.

The {article} is organized as follows: the methodology is discussed in Section~\ref{sec:method}. 
The application of the method for four prototypical transition-metal oxides is presented in Section~\ref{sec:results}, and the results are compared against available experimental and theoretical data. 
Section~\ref{sec:discussion} discusses the important features of the method and suggests extensions of potential significance to the goal of discovering novel functional materials.
 {Conclusions are summarized in Section~\ref{sec:conclusions}.}

\section{Methodology} \label{sec:method}

The foundations of the approach for evaluating the on-site Coulomb and exchange parameters are:

\textit{i}) $\bar{U}$ and $\bar{J}$ are on-site quantities derived from the energy in the Hartree-Fock method.
The HF theory considers pair-wise interactions of only two electrons at the time and therefore it misses the concept of screening. Because of this approximation, the HF theory is known to be inappropriate in describing delocalized metallic systems; however, it is qualitatively sound for molecules and insulators (especially in the strong localization regime)\cite{orlando1990ab,dovesi1992ab,dovesi1990abinitio,dovesi2000periodic}. 
The on-site energies are derived from the HF energies following the ansatz of Mosey and Carter \cite{Mosey2007,Mosey2008} in which the occupied molecular orbitals (MO), needed for the computation of the HF energy, are considered populated only in the subspace $\{m\}$.

\textit{ii}) no localized orbitals $\varphi_{i}$ need to be explicitly computed. As in the Hartree-Fock method, all the MOs, or crystalline  wavefunctions for the case of solids, are used. This eliminates the indeterminacy in finding the subset of MOs that better corresponds to the localized states, which can lead to a wide fluctuations of the calculated $\bar{U}$ \cite{Andriotis_LSDAU_HF_PRB2010}. During the calculation of the on-site HF energies, the localized orbitals are implicitly taken as a linear combinations the basis functions of interest, $\{m\}$, with the expansion coefficients included in the renormalized density matrix coming directly from the solution of the Kohn-Sham equations projected onto the localized basis of choice (see below); 

\textit{iii}) {a plane-wave basis set is the natural choice for DFT calculations of periodic systems, but on-site HF energies are more efficiently computed in a localized basis set.}
 {Electron-repulsion integrals are evaluated} using pseudo-atomic-orbitals (PAO) expressed as linear combination of Gaussian-type functions, that we define as the PAO-3G minimal basis set.
This is possible by the projection procedure that we have recently developed \cite{curtarolo:art86}, which seamlessly maps the plane-wave electronic structure onto a localized atomic-orbital basis set (see Appendix A). However, it is important to note that the construction of $\bar{U}$ and $\bar{J}$ outlined below is completely general and can be applied to any choice of basis, localized or otherwise.

\textit{iv}) $E^{\{m\}}_{\textrm{HF}}$ is a true functional of the electron density in the spirit of the Hohenberg-Kohn theorems. 
{This leads to the definition of the \ACBN\, pseudo-hybrid Hubbard density functional.}

\subsection{Calculation of the Electron Repulsion Integrals} \label{subsec:eri}

The enormous quantity of {Electron Repulsion Integrals, ERIs}, needed in the calculation of the HF exchange energy is the fundamental bottleneck in the use of  hybrid DFT functionals.
In DFT calculations based on LCAO (PAO) basis sets, the problem is made more tractable when the PAOs are expressed as linear combinations of Gaussian-type functions, as it is commonly done in commercial packages such as Gaussian09 \cite{Gaussian_2009} and Crystal06 \cite{Crystal_2006}. 

The electron repulsion integrals used in Eq.~\ref{eq:liechtenstein}  are defined as four PAOs interacting under  the bare Coulomb interaction $V=\mathbf{|r_{1}-r_{2}|}^{-1}$ as:
\begin{align} \label{eq:two_e}
\textrm{ERI} & {\equiv (mm'|m''m''') \notag} \equiv \langle mm''|V|m'm''' \rangle\\ 
& \equiv \int  d\mathbf{r_1} d\mathbf{r_2}\phi_{m}^{*}(\mathbf{r_1})\phi_{m'}(\mathbf{r_1}) V \phi_{m''}^{*}(\mathbf{r_2})\phi_{m'''}(\mathbf{r_2}).
\end{align}



The real-space evaluation of these integrals is not directly possible when using a plane-wave basis set, which is the preferred choice for periodic systems. 
For this reason, we employ the auxiliary space of {PAOs} 
naturally included in the definition of the pseudo-potentials. 
Given that the radial and angular part of the PAO basis functions, $\phi_{lm}(\mathbf{r}) \equiv \frac{R_{l}(r)}{r} Y^{m\{c,s\}}_{l}(\theta,\varphi)$,  are separable, they can be directly fitted using linear combinations of spherical-harmonic Gaussian functions.
For efficiency, the {latter} functions are then further expanded as linear combination of Cartesian Gaussians defining the PAO-3G minimal basis set.
 (see Appendix~\ref{sec:fitting} for more technical details on these transformations).
 Once expressed in the PAO-3G basis set, the ERIs can be efficiently evaluated using any optimized quantum-chemistry library. 
We use the C routines included in the open-source quantum-chemistry package PyQuante  
\footnote{\label{note1} Richard P. Muller, PyQuante 1.6.4: Python Quantum Chemistry \texttt{http://pyquante.sourceforge.net/}}.




\subsection{Hartree-Fock Coulomb and exchange energies} 
The knowledge of the ERIs and the molecular (or crystal) orbitals allows the calculation of the HF Coulomb and exchange energies $E_{\textrm{HF}}$. For isolated systems (molecules or clusters) and in the restricted case: 

\begin{align} \label{eq:HF}
E^{\textrm{molec.}}_{\textrm{HF}}&= \sum_{ij}   N_{\psi_i} N_{\psi_j} 
\left[ 2(\psi_i \psi_i | \psi_j \psi_j) - (\psi_i \psi_j | \psi_j \psi_i) \right] \nonumber \\ 
&= \sum_{\mu\nu\kappa\lambda} P_{\mu\nu}P_{\kappa\lambda}[2(\mu \nu | \kappa \lambda) - (\mu\lambda|\kappa\nu)]. 
\end{align}
Here $\psi^{\sigma}_i(\mathbf{r})=\sum_{i,\mu} c^{\sigma}_{\mu i} \phi_{\mu}(\mathbf{r})$ are occupied molecular orbitals expanded in the {PAOs' basis}; $N^{\sigma}_{\psi_i} \equiv \sum_{i\mu\nu}c^{\sigma *}_{\mu i}S_{\mu\nu}c^{\sigma}_{\nu i} = 1$ is the charge of $\psi^{\sigma}_i$ and $S_{\mu\nu}$ is the overlap integral between the PAOs $\phi_{\mu}$ and $\phi_{\nu}$.
The last line of Eq.~\ref{eq:HF} is expressed in the basis of atomic orbitals $\phi_{\mu}$ with the density matrix
$P^{\sigma}_{\mu\nu}=\sum_i N^{\sigma}_{\psi_i} c^{\sigma *}_{\mu i}c^{\sigma}_{\nu i}$. 

The expression of the Coulomb and exchange HF energies for a periodic system is analogous to the molecular case (see Pisani \textit{et al.} \cite{Pisani_HF_exact_periodic_IJQC1980}):  
\begin{equation} \label{eq:HFsolid}
E^{\textrm{solid}}_{\textrm{HF}}\!=\!\!\!
\sum_{\mu\nu\kappa\lambda \atop \mathbf{g,l,m}}\!\!P^{\mathbf{g}}_{\mu\nu} P^{\mathbf{l}}_{\kappa\lambda}\left[2\left({\mu}^{\mathbf{0}} {\nu}^{\mathbf{g}} | {\kappa}^{\mathbf{m}}{\lambda}^{\mathbf{m+l}}\right) - \left({\mu}^{\mathbf{0}} {\kappa}^{\mathbf{m}}|{\nu}^{\mathbf{g}}{\lambda}^{\mathbf{m+l}}\right)\right],
\end{equation}
where $\mathbf{g}$, $\mathbf{m}$ and $\mathbf{l}$ are lattice vectors and $\mathbf{0}$ refers to the primitive unit cell.
However, the mapping of the crystalline wavefunctions $\psi^{\mathbf{k}\sigma}_i$ in a local basis (i.e. the expansion coefficients $c^{\mathbf{k}\sigma}_{\mu i}$) is not readily available when using a plane-wave basis to solve for the electronic structure of the material as it is common for solids. We circumvent this problem by projecting the plane-wave solution into the chosen auxiliary space of PAOs following the method described in Ref.~[\onlinecite{curtarolo:art86}]. This projection procedure is a noniterative scheme to represent the electronic ground state of a periodic system using an atomic-orbital basis, up to a predictable number of electronic states, and with controllable accuracy by filtering out high-kinetic-energy plane waves components. See Appendix~\ref{sec:hm_dm} for a summary of this procedure to calculate the expansion coefficients $c^{\mathbf{k}\sigma}_{\mu i}$ and the real-space density matrices of the solid, $P^{\sigma,\mathbf{R}}_{\mu\nu}$.

\subsection{ $\bar{U}$ and $\bar{J}$ as functional of the density: the \ACBN\, functional} \label{subsec:reduced_HF_energies}

The energy functional for the DFT+$U$ method is given by:
\begin{equation*}
E_{\textrm{DFT}+U}=E_{\textrm{DFT}} + E_{U}
\end{equation*} 
where $E_{\textrm{DFT}}$ is the DFT energy calculated using a LDA or GGA functional. The energy correction $E_U$ is given either in the original Anisimov-Liechtenstein \cite{Liechtenstein1995,Anisimov1997} or in the simplified Dudarev\cite{Dudarev1998} formulation as:
\begin{subequations} 
\begin{gather}
E^{\textrm{Anisimov}}_{U} =\left[ \sum\limits_{I} E^{\{m\},I}_{\textrm{HF}} \right] - E_{\textrm{DC}}, \\ 
E^{\textrm{Dudarev}}_{U}  =\frac{\bar{U}-\bar{J}}{2}\sum\limits_{I}\sum_{m,\sigma}\left[ n^{I\sigma}_{mm} -\sum\limits_{m'}n^{I\sigma}_{mm'} n^{I\sigma}_{m'm} \right], \label{eq:E_U_dudarev}
\end{gather}
\end{subequations}
with $E^{\{m\},I}_{\textrm{HF}}$ defined in Eq.~\ref{eq:liechtenstein} for a given atom $I$. $E_{\textrm{DC}}$ corrects for a possible double counting of the localized-states interaction energy already captured (in an averaged way) in $E_{\textrm{DFT}}$. The second formulation defines an effective on-site Coulomb interaction $U_{\textrm{eff}}=\bar{U}-\bar{J}$ (henceforth referred simply as $U$). It should be noticed that numerical implementations of the Anisimov DFT+$U$ functional (Eq.~\ref{eq:liechtenstein}), for instance in \textsc{quantum espresso} \cite{quantum_espresso_2009} or VASP \cite{vasp}, do not compute the ERIs explicitly. They are evaluated from tabulated Slater integrals, which ultimately depend on the provided values of $\bar{U}$ and $\bar{J}$, {or from phenomenological considerations (e.g. Ref.~[\onlinecite{aflowBZ,aflowTHERMO}]).}

On the contrary, we evaluate $\bar{U}$ and $\bar{J}$ by computing directly the on-site Coulomb and exchange energies on the chosen Hubbard center, from the Coulomb and exchange Hartree-Fock energies of the solid. The following assumptions are used.

(\textit{i}) We follow a central ansatz, introduced by Mosey \textit{et al.} \cite{Mosey2007,Mosey2008} for the case of cluster calculations, that defines a ``renormalized'' occupation number $\overline{N}^{\sigma}_{\psi_i} \neq 1$ for each MO $\psi^{\sigma}_i$:
\begin{equation} \label{eq:focc}
\overline{N}^{\sigma}_{\psi_i} \equiv \sum_{\mu \in \{\overline{m}\}}\sum_{\nu}c^{\sigma *}_{\mu i}S_{\mu\nu}c^{\sigma}_{\nu i},
\end{equation}
which is the Mulliken charge of the basis $\{\overline{m}\}$. The set $\{\overline{m}\}$ includes all the atomic orbitals in the unit cell that have the same quantum numbers as the orbitals $\{m\}$ of the Hubbard center of interest.

Correspondingly, we define a renormalized density matrix as: 
\begin{equation} \label{eq:fdm}
\bar{P}^{\sigma}_{\mu\nu} \equiv \sum_i \overline{N}^{\sigma}_{\psi_i} c^{\sigma *}_{\mu i}c^{\sigma}_{\nu i}.
\end{equation}
The renormalized occupations can be interpreted as weighting factors that specify the on-site occupation of each electronic state.

The expressions in Eqs.~\ref{eq:focc} and \ref{eq:fdm} are applicable to isolated systems (molecules and clusters). Solid-state systems are calculated in reciprocal $\mathbf{k}$-space that inherently apply periodic-boundary conditions, thus, surface effects are avoided. Periodic plane waves are the basis of choice for the solution of solid-state systems. The $\mathbf{k}$-space electronic-structure information of the material can be projected into a PAO basis directly from the plane-wave DFT solution. We take advantage of this to compute the reduced density matrices and occupation number, needed to calculate $\bar{U}$ and $\bar{J}$, as follows
\begin{subequations}
\label{eq:rfocc} 
\begin{gather}
\bar{P}^{\sigma}_{\mu\nu}=\bar{P}^{\mathbf{0},\sigma}_{\mu\nu}=\frac{1}{\sqrt{N_{\mathbf{k}}}} \sum\limits_{\mathbf{k},i}  
\overline{N}^{\mathbf{k}\sigma}_{\psi_i}
c^{\mathbf{k}\sigma *}_{\mu i}c^{\mathbf{k} \sigma}_{\nu i}
\\
\overline{N}^{\mathbf{k}\sigma}_{\psi_i}=\sum_{\kappa \in \{\overline{m}\}} \sum_{\lambda}c^{\mathbf{k}\sigma *}_{\kappa i}S^{\mathbf{k}}_{\kappa \lambda}c^{\mathbf{k} \sigma}_{\lambda i}  \label{eq:kfocc}
\\
N^{\sigma}_{m} =\frac{1}{\sqrt{N_{\mathbf{k}}}} 
\sum\limits_{\mathbf{k},i,\nu}  
c^{\mathbf{k}\sigma *}_{m i}S^{\mathbf{k}}_{m\nu}c^{\mathbf{k}\sigma}_{\nu i}.
\end{gather}
\end{subequations}
Here, $N_\mathbf{k}$ is the total number of $\mathbf{k}$-vectors in the $1^{st}$ Brillouin {zone}. See Appendix~\ref{sec:hm_dm} for more details.  

(\textit{ii}) $E^{\{m\}}_{\textrm{HF}}$, the on-site HF energy associated to the basis $\{m\}$, is obtained from Eq.~\ref{eq:HF} by restricting the summation indexes to $\{m\}$. In the periodic case, it is reduced from Eq.~\ref{eq:HFsolid} considering the central unit cell only, i.e. lattice vectors $\mathbf{R=g=l=m=0}$. Combining (\textit{i}) and (\textit{ii}), we obtain in the general spin-unrestricted case:
\begin{align} \label{eq:HF3}
E^{\{m\}}_{\textrm{HF}}&=\frac{1}{2} \sum_{\{m\}} [ \bar{P}^{\alpha}_{mm'}\bar{P}^{\alpha}_{m''m'''} + \bar{P}^{\alpha}_{mm'}\bar{P}^{\beta}_{m''m'''} \nonumber \\
&+ \bar{P}^{\beta}_{mm'}\bar{P}^{\alpha}_{m''m'''}+ \bar{P}^{\beta}_{mm'}\bar{P}^{\beta}_{m''m'''}] (mm'|m''m''') \nonumber \\ 
&+\frac{1}{2} \sum_{\{m\}} [ \bar{P}^{\alpha}_{mm'}\bar{P}^{\alpha}_{m''m'''} + \bar{P}^{\beta}_{mm'}\bar{P}^{\beta}_{m''m'''}] (mm'''|m''m'). 
\end{align}

In this notation, each primed and unprimed index $m$ runs over all the set $\{m\}$. Clearly, the above equation is equivalent to  Anisimov's original DFT+$U$ functional (Eq.~\ref{eq:liechtenstein}) once we replace $n^{\sigma}_{mm'}$ with the  renormalized density matrix $\bar{P}^{\sigma}_{mm'}$. $\sigma=\{\alpha,\beta\}$. However, while Eq.~\ref{eq:liechtenstein} requires the knowledge of a {subjective} screened Coulomb interaction $V_{ee}$, Eq.~\ref{eq:HF3} uses the bare Coulomb interaction. Although screening is not considered in the HF theory, the direct parallel between both equations shows that the renormalization of the density matrix effectively introduces a degree of screening.

 {The comparison of Eqs.~\ref{eq:dudarev} with \ref{eq:HF3} leads to the definitions of} $\bar{U}$ and $\bar{J}$ as density dependent quantities in the \ACBN\, functional:  
\begin{widetext}
\begin{equation} \label{eq:param_u}
\bar{U}=\frac{\sum_{\{m\}} [ \bar{P}^{\alpha}_{mm'}\bar{P}^{\alpha}_{m''m'''} + \bar{P}^{\alpha}_{mm'}\bar{P}^{\beta}_{m''m'''} + \bar{P}^{\beta}_{mm'}\bar{P}^{\alpha}_{m''m'''}+ \bar{P}^{\beta}_{mm'}\bar{P}^{\beta}_{m''m'''}](mm'|m''m''')}
{\sum_{m \neq m'}N^{\alpha}_{m}N^{\alpha}_{m'} + \sum_{\{m\}}N^{\alpha}_{m} N^{\beta}_{m'} + \sum_{\{m\}}N^{\beta}_{m} N^{\alpha}_{m'} + \sum_{m \neq m'}N^{\beta}_{m}N^{\beta}_{m'}},
\end{equation}

\begin{equation} \label{eq:param_j}
\bar{J}=\frac{\sum_{\{m\}} [ \bar{P}^{\alpha}_{mm'}\bar{P}^{\alpha}_{m''m'''} + \bar{P}^{\beta}_{mm'}\bar{P}^{\beta}_{m''m'''}](mm'''|m''m')}
{\sum_{m \neq m'}N^{\alpha}_{m}N^{\alpha}_{m'} + \sum_{m \neq m'}N^{\beta}_{m}N^{\beta}_{m'}}.
\end{equation}
\end{widetext}


 {There are essential and relevant differences between Eqs.~\ref{eq:param_u} and \ref{eq:param_j} and the similar framework of Mosey \textit{et al.} \cite{Mosey2008}:}
\textit{i}) our on-site energy requires the computation of a smaller number of electron-repulsion integrals, namely only those involving the $\{m\}$ set (i.e. $5^4$ integrals for the $d$ shell). This directly parallels the original definition of Anisimov, in sharp contrast with the methodology of Mosey \textit{et al.} requiring ERIs between all basis functions contained in the cluster; \textit{ii}) we consider the larger set $\{\overline{m}\}$ of localized orbitals (instead of $\{m\}$) in the calculation of the reduced occupation number given in Eq.~\ref{eq:focc}; \textit{iii}) in the Mosey \textit{et al.} \cite{Mosey2008} approach the atom of interest is embedded in a cluster of volume large enough to yield local convergence to bulk conditions, thus requiring calculations on clusters of increasing volume to ensure convergence, which can become extremely computationally expensive; \textit{iv}) finally, we never solve the full Hartree-Fock problem for the solid (Roothaan's equations) but project the DFT Kohn-Sham wavefunctions on the minimal PAO-3G basis set, thus implicitly including in the renormalized density matrix all the local screening effects that come from the mean field solution on the local set.

\section{Results}\label{sec:results}
We selected four prototypical examples to benchmark the \ACBN\, density functional:
TiO$_2$ (rutile), MnO, NiO, and ZnO (wurtzite) are technological important  transition metal oxides (TMO) that have been extensively studied both 
theoretically and experimentally. These materials pose a methodological challenge to traditional energy functionals (LDA or GGA) due to the strong localization 
of the TM-$3d$ electrons that results in significant errors in the description of their electronic {structure \cite{aflowlibPAPER}.}
The set of chosen TMOs covers a wide range of $3d$ shell fillings, namely, $3d^2$, $3d^5$, $3d^8$, and $3d^{10}$ in Ti, Mn, Ni and Zn, respectively. 

All our calculations use the  Perdew-Burke-Ernzerhoff (PBE) \cite{PBE} functional as a starting point and a plane wave energy cutoff of 350 Ry with a dense Monkhorst-Pack mesh to ensure good convergence of all quantities. All DFT+$U$ calculations use the simplified rotational-invariant scheme of Dudarev \cite{Dudarev1998} and Cococcioni \cite{Cococcioni_DFTU_LinearResponse_PRB2005} as implemented in the \textsc{quantum espresso} package \cite{quantum_espresso_2009}. It is noticed that both the original DFT+$U$ formulation of Anisimov and the simplified rotationally-invariant scheme are equivalent within numerical error for the same values of $\bar{U}$ and $\bar{J}$. 
For all elements, we used scalar-relativistic norm-conserving pseudopotential from the PSlibrary 1.0.0.\footnotemark[1]

Although the calculation of the effective values of $\bar{U}$ and $\bar{J}$ should be performed {concurrently} within the general Kohn-Sham 
self-consistent loop for electronic convergence, in this work we follow a simplified scheme: the initial {and objective} guess for the first DFT+$U$ calculation is 
 {$U^{(0)}_{3d}\!=\!U^{(0)}_{2p}\!=\!0$ eV}.
 Then, the resulting electronic structure is used to compute $U^{(1)}$ for the next DFT+$U$ step (from Eqs.~\ref{eq:param_u} and \ref{eq:param_j}). The process is iterated simultaneously for both transition metal and oxygen atoms until the difference between  two subsequent iterations is {$\left|U^{(n)}-U^{(n-1)}\right|< 10^{-4}$} eV. This self-consistent scheme ensures the internal consistency of the results while the true variational solution using the \ACBN\, functional will be implemented in the near future. 
The converged effective values of $U$ for the transition-metal oxides under study are reported in Table~\ref{tab:ueff}.  
  All the presented band structures follow the {\small AFLOW} standard integration paths \cite{aflowBZ}.
  
In what follows, the ACBN0 results will be benchmarked against experimental measurements and higher level of theory, whenever possible. Clearly, meaningful comparisons between theory and photoemission spectra require $GW$ quasiparticle energies (and more for excitonic effects) which go beyond DFT. Even if the DFT+$U$ formalism has been shown to be a first-order approximation to the $GW$ method for localized states in the static limit,\cite{Anisimov1997,Jiang_localized_GWLDAU_PRB2010} in practice ACBN0 is clearly not nearly as inclusive and accurate as  $GW$  and a comparison between the results of the two approaches provides with a most stringent test of the validity of our method.

\subsection{Titanium dioxide (rutile)}
\begin{figure*}[th]
  \includegraphics[width=0.99\textwidth]{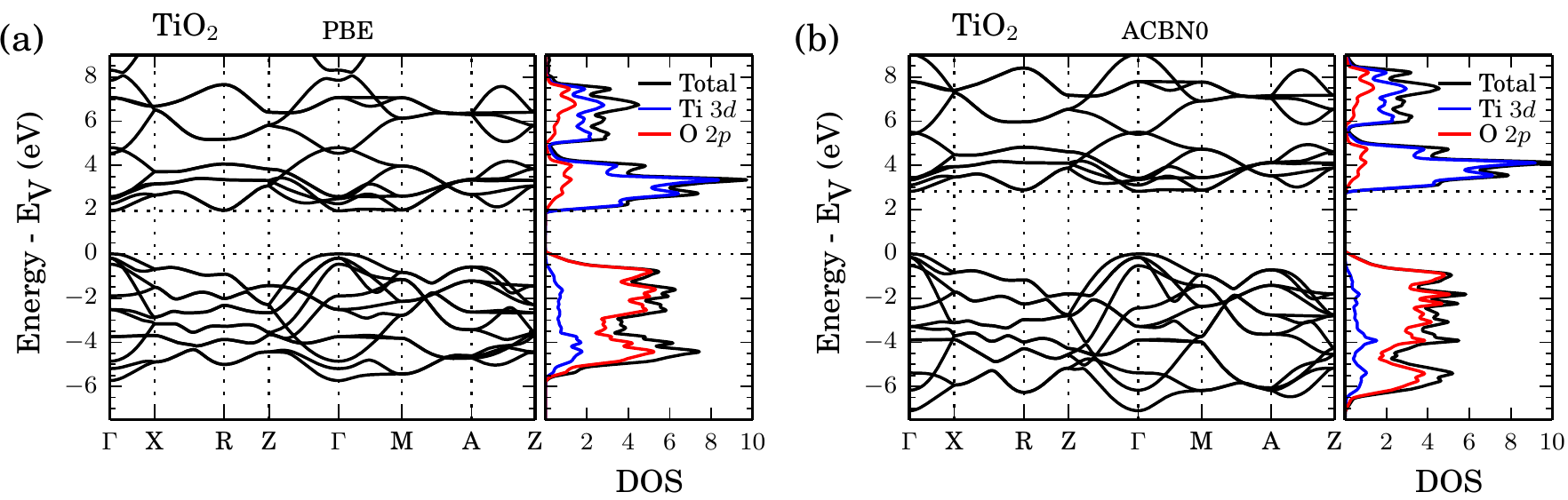}
\vspace{-3mm}
  \caption{\small
    Comparison between the band structures and projected density of states of TiO$_2$ without
    on-site interactions $U=0$ (a) and with the converged effective values of $U$ = 0.15 eV for Ti and 7.34 for O (b). } 
  \label{fig:fig_tio2}
\end{figure*}
Rutile, with space group \textit{P}4$_{2}/mnm$ ( {\#}136) is the most common form of TiO$_2$. We use the experimental lattice constants and internal ordering parameter of $a=b=4.594$~\AA, $c=2.959$~\AA, $\mu=0.305$ \cite{Abrahams_JCP1971}

The valence manifold is predominantly of O-$2p$ character with small Ti-$3d$ hybrization except at the top of the manifold at $\Gamma$, where it takes almost exclusively an O-$2p$ character. Conversely, the unoccupied manifold is predominantly of Ti-${3d}$ character; the conduction-band minimum (CBM) is at $\Gamma$ but in practice it is degenerate with the minima at R and M. Two regions are distinguishable in the $3d$ projected density of states (PDOS) in the unoccupied manifold (Fig.~\ref{fig:fig_tio2}) and correspond to the octahedral-type crystal-field splitting of $e_g$ (higher energy) and $t_{2g}$ states (lower energy). 

The use of an increasing on-site Coulomb potential $U_{3d}$ on Ti alone (without correcting the oxygen) has been shown to monotonically open the gap, which reaches satisfactory accord with the experimental value of 3.03 eV \cite{Pascual_TiO2_PRB1978} only at values of $U_{3d}\sim 10$ eV \cite{Persson_polaron_TiO2_APL2005,Deskins_polaron_hopping_TiO2_PRB2007}. However, Park \textit{et al.} \cite{Park_LDAU_TiO2_PRB2010} have found that large values of the Ti on-site Coulomb interaction ($>7$) introduces unphysical defect states in the study of vacancies and suggested a concomitant use of $U_{2p}=7$ eV on oxygen is necessary to achieve both the experimental bandgap and a good treatment of vacancy states.

With small L\"{o}wdin charges of 0.29$e$--0.45$e$ (out of 2$e$) per orbital \cite{Lowdin:1970:NP}, the Ti-$3d$ states can not be considered localized and therefore the use of large values of $U_{3d}$ is understood as an \textit{ad hoc} fitting parameter without physical basis. Instead, each oxygen $2p$ orbital charge is 1.66$e$ (out of 2$e$).

Our converged values for the rutile environment are Ti $U_{3d}=0.15$ eV and O $U_{2p}=7.34$ eV. These values yield a bandgap of 2.83 eV close to the experimental range of 2.8--3.8 eV (Table~\ref{tab:gaps}), which improves the DFT prediction by 0.9 eV. Contrary to the predominant focus on the Ti-$3d$ states, our results show that a correction on the  oxygen $2p$ states can alone yield an equally satisfactory bandgap. More generally, it suggests that a correct treatment of oxygen $2p$ states may be more relevant to the correct modeling of TiO$_2$ vacancies.

The occupied O-$2p$ PDOS [red line in Fig.~\ref{fig:fig_tio2}(a)] shows a split into two regions, upper 0--3 eV and lower 3.5--6 eV, which originates in the oxygen $2p$ crystal-field splitting. The $2p_x$ and $2p_y$ states form $sp^2$-like $\sigma$ bonds contained in the planar Y-shaped OTi$_3$ subunits whereas the $2p_z$ states remain as lone pairs perpendicular to the Y-shaped planes. The higher-energy $2p_z$ states correspond to the upper PDOS region. These nonbonding lone pairs have been explained with a simple empirical molecular-orbital model\cite{Grunes_oxygen_fine_structure_TMO_PRB1982, Woicik_hybridization_bond_orbital_TiO2_PRL2002} whereby the octahedral $O_h$ symmetry of the local environment of each Ti coordinated to six oxygen ligands (TiO$_6$)$^{8-}$ frustrates the hybridization of the highest occupied orbitals\cite{Ballhausen_molecular_orbital_book_1965}.

Applying the on-site Coulomb potential $U_{2p}$ on oxygen increases the localization of the $2p_z$ lone pairs, thus, increasing the splitting between the two $2p$ PDOS regions, cf. Figs.~\ref{fig:fig_tio2}(a) and (b). The main peak of the lower PDOS region downshifts by 1 eV, to $\sim$5.4 eV, consistent with the value of 5 eV reported by the accurate {full-frequency-dependent } $GW$ calculation of Khan and Hybertsen $GW$ \cite{Kang_quasiparticle_TiO2_PRB2010} and X-ray photoelectron spectroscopy (XPS) measurements \cite{Kowalczyk_elecronic_structures_oxides_SSC1977}.  

Examining the $G_0W_0$@GGA band structure reported by Malashevich \textit{et al.} \cite{Malashevich_GW_TiO_PRB2014}, it is interesting to notice that besides a scissor-shift operation the main correction with respect to the DFT bands is a downward energy shift of the $2p_x2p_y$ bands (lower region of the occupied manifold, -6 to -4 eV) whereas the upper region (-4 to 0 eV) remains mostly unchanged. A possible mechanism is that the $GW$ approach implicitly applies a self-interaction correction that increases the splitting between the $2p_z$ and $2p_x2p_y$ states by further localizing the $2p_z$ states, which can be captured in the \ACBN\, calculation.

In this regard, \ACBN\, bands closely follow the $G_0W_0$@GGA bands of Ref.~\onlinecite{Malashevich_GW_TiO_PRB2014} in the range from -4 to 8 eV. The most significant difference happens in the remaining range of -7 to -4 eV, where our DFT+$U$ bands are over downshifted with respect to the $G_0W_0$@GGA results. This can be expected from the explicit emphasis of on-site Coulomb interaction on oxygen in the \ACBN\, approach.

\begin{table}
\caption{\label{tab:ueff} Converged values of the effective on-site Coulomb parameter $U$ (in eV) for the transition metal (TM) $3d$ and the oxygen $2p$ states.}
\begin{ruledtabular}
\begin{tabular}{lcccc}
   & TiO$_2$  &MnO & NiO & ZnO \\
\hline
TM-$3d$      & 0.15 & 4.67  & 7.63 & 12.8 \\
Oxygen $2p$  & 7.34 & 2.68  & 3.0  & 5.29 \\
\end{tabular}
\end{ruledtabular}
\end{table}

\begin{table*}
\caption{\label{tab:gaps} Minimum direct and indirect energy bandgaps (in eV).}
\begin{ruledtabular}
\begin{tabular}{lcccccc}
 &TiO$_2$ & \multicolumn{2}{c}{MnO} & \multicolumn{2}{c}{NiO} & ZnO\\
 &dir.    &indir.&dir.              &indir.&dir.              & dir.\\
\hline
PBE     & 1.94 & 0.98  & 1.64 & 1.13   & 1.26   & 0.85 \\
\rowcolor{Gray}
\ACBN\, & 2.83 & 2.31  & 2.83 & 3.80   & 4.29   & 2.91 \\
sX-LDA         & 3.1 \cite{Lee_oxygen_TiO2_MRS2011} & 2.5 \cite{Gillen_TMO_JPCM2013} & 3.0 \cite{Gillen_TMO_JPCM2013} & 4.04 \cite{Gillen_TMO_JPCM2013} & 4.3 \cite{Gillen_TMO_JPCM2013} &  3.1 \cite{Akiyama_JPCS2008}, 3.41 \cite{Clark_screened_exchange_PRB2010} \\
HSE03          & 3.25 \cite{Nakai_HSE_TiO2_JAPAN2006}  & 2.6 \cite{Rodl_quasiparticles_antiferro_oxides_PRB2009} & 3.2 \cite{Rodl_quasiparticles_antiferro_oxides_PRB2009} & 4.1 \cite{Rodl_quasiparticles_antiferro_oxides_PRB2009} & 4.5 \cite{Rodl_quasiparticles_antiferro_oxides_PRB2009} & 2.11 \cite{Fuchs_quasiparticles_KS_PRB2007} \\
$G_0W_0$@GGA   & 3.18 \cite{Malashevich_GW_TiO_PRB2014}, 3.4 \cite{Patrick_GW_TiO2_JPCM2012}  & 1.7 \cite{Rodl_quasiparticles_antiferro_oxides_PRB2009} & 2.1 \cite{Rodl_quasiparticles_antiferro_oxides_PRB2009} & 1.1 \cite{Rodl_quasiparticles_antiferro_oxides_PRB2009} & 1.4 \cite{Rodl_quasiparticles_antiferro_oxides_PRB2009} & 2.12 \cite{Shishkin_scGW_PRB2007}\\
$G_0W_0$@HSE03 &3.73 \cite{Landmann_electronic_structure_response_TiO2_JPCM2012}\footnote{@HSE06}  & 3.4 \cite{Rodl_quasiparticles_antiferro_oxides_PRB2009} & 4.0 \cite{Rodl_quasiparticles_antiferro_oxides_PRB2009} & 4.7 \cite{Rodl_quasiparticles_antiferro_oxides_PRB2009} & 5.2 \cite{Rodl_quasiparticles_antiferro_oxides_PRB2009} & 2.97 \cite{Fuchs_quasiparticles_KS_PRB2007}\\
$GW$@\{LDA/GGA\}       &  & 3.5 \cite{Faleev_GW_Si_MnO_NiO_PRL2004} & & 4.8 \cite{Faleev_GW_Si_MnO_NiO_PRL2004} & & 2.92 \cite{Lim_arpes_calcu_ZnO_PRB2012}, 3.2 \cite{Shishkin_scGW_PRB2007}\\
 
Exp. (XAS-XES) &      & \multicolumn{2}{c}{4.1 \cite{Kurmaev_correlated_oxides_PRB2008}} & \multicolumn{2}{c}{4.0 \cite{Kurmaev_correlated_oxides_PRB2008}} &  3.3 \cite{Dong_ZnO_PRB_2004} \\
Exp. (PES-BIS) & 3.3$\pm$0.5 \cite{Tezuka_photoemission_bremsstralug_TiO2_SrTiO3_JAPAN1994} & \multicolumn{2}{c}{3.9 $\pm$ 0.4 \cite{Elp_MnO_PRB1991}}     & \multicolumn{2}{c}{4.3 \cite{Sawatzky_gapNiO_PRL1984}}     &       \\ 
Exp. (Conductance.)&      & \multicolumn{2}{c}{3.8--4.2 \cite{Drabkin_FIz1968}}    & \multicolumn{2}{c}{3.7 \cite{Ksendzov_FIZ1965}} & \\
Exp. (absorption)       & 3.03 \cite{Pascual_TiO2_PRB1978}  & \multicolumn{2}{c}{3.6--3.8 \cite{Iskenderov_FIZ1968}} & \multicolumn{2}{c}{} & 3.44 \cite{Mang_ZnO_SSC1995}\\
Exp. (reflectance)       &      &  \multicolumn{2}{c}{} & \multicolumn{2}{c}{3.7 \cite{Powell_optical_NiO_CoO_PRB1970}, 3.87 \cite{Kang_NiO_CoO_ellipsometry_Korean2007}} & 3.44 \cite{Reynolds_valence_band_ZnO_PRB1999}\\
\end{tabular}
\end{ruledtabular}
\end{table*}

\subsection{Manganese and nickel oxides}

For MnO (NiO), we use the ideal rocksalt structure with lattice constant $a=4.4315$~\AA\ ($a=4.1704$~\AA) \cite{Cheetham_magnetic_ordering_MnNiO_PRB1983}. The presence of type-II antiferromagnetic spin coupling along the [111] direction, below N\'{e}el temperature, effectively requires a rhombohedral primitive unit cell (RHL$_1$ \cite{aflowBZ}, $a_{\textrm{RHL}}=a\sqrt{{3}/{2}}, \alpha=33.557^{\circ}$) containing 4 atoms with space group $R\bar{3}m$ ( {\#}166).

Our converged values for MnO are $U_{3d}=4.67$ eV for Mn and $U_{2p}=2.68$ eV for O. This value is in the range of other \textit{ab-initio} $U$s reported for Mn (3.6--6.04 eV) \cite{Pickett1998,Jiang_localized_GWLDAU_PRB2010,Floris_vibration_MnO_NiO_DFTU_PRB2011,Shih_screened_interaction_TMO_PRB2012,Anisimov_Mott_insulators_JPCM1990}. It should be noticed that due to the different assumptions for the physical quantities (i.e. screening, localized states), \textit{ab-initio} values of $U$ should not be expected to the unique. A close empirical value of $U_{3d}=4.0$ eV (albeit with no correction on oxygen) has been reported to reproduce well the experimental energy of formations of several manganese oxides \cite{Wang_Ceder_GGAU_PRB_2006,Franchini_manganese_oxides_PRB2007}. 

\begin{figure}[htb]
\begin{center}
    \includegraphics[width=0.99\columnwidth]{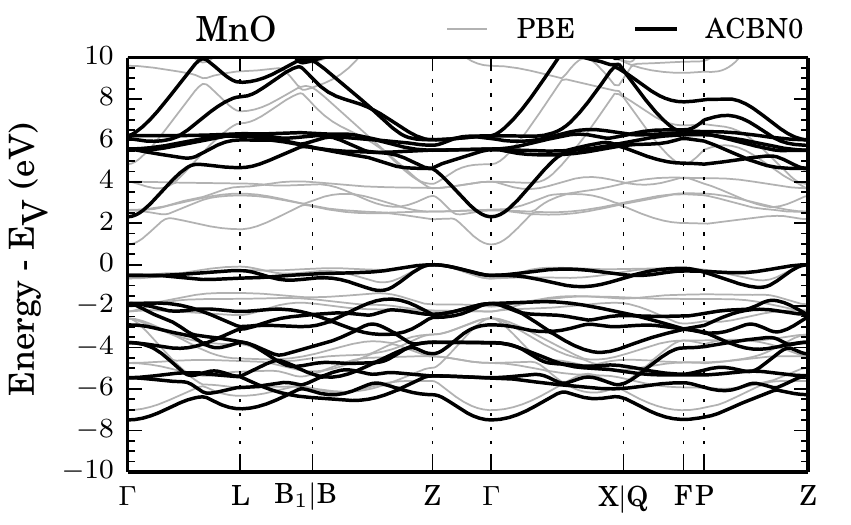}
    \vspace{-5mm}
    \caption{\small
      Band structure (spin up) of manganese oxide. All energies are relative to the valence band maximum E$_\textrm{V}$. Effective values of $U=4.67$ eV for the {Mn-$3d$}
      states and 2.68 eV for the {O-$2p$} states are used in the \ACBN\ calculation.}
    \label{fig:fig_mno} 
  \end{center}
\end{figure}

Both PBE and \ACBN\, band structures are shown in Fig.~\ref{fig:fig_mno}. The bottom of the conduction manifold is an itinerant $sp$ band with noticeable parabolic dispersion and is predominantly of Mn-$4s$ character at CBM at $\Gamma$. The set of low-dispersion bands located above the CBM are predominantly Mn-$3d$ $t_{2g}$ states. These bands are more narrowly resolved in the case of NiO.

In the occupied manifold, the PBE results (gray lines) distinctly show dispersionless Mn $e_g$ bands (separated from the rest at the top of the manifold) centered at $\sim -0.5$ eV and $t_{2g}$ bands at $\sim -1.5$ eV. These bands have minor oxygen hybridization whereas the bands below them ($\lesssim -1.6$ eV) have a strong O-$2p$ {character}.
With the on-site repulsion correction in the \ACBN\, results (black lines), the hybridization between Mn-$3d$ and O-$2p$ states increases. As a result the $3d$ bands are pushed down in energy {while increasing} their dispersion (more noticeably on the $t_{2g}$ bands). This leads to:
\textit{i}) an increase of the bandwidth of the occupied manifold to $\sim 7.5$ eV, in good agreement with results from higher levels of theory (7.6 eV $GW$@LDA+$U$\cite{Kobayashi_GWLDAU_PRB2008} and 8 eV sX-LDA \cite{Gillen_TMO_JPCM2013}), 
\textit{ii}) an increase of the $t_{2g}$ bonding-antibonding splitting across the bandgap and indirectly the resolution of the $4s$ parabolic band (i.e. the energy difference between the parabolic CBM at $\Gamma$ and the occupied $t_{2e}$ bands) to 2.21 eV, which compares favorably with the more rigorous self-consistent $GW$ \cite{Faleev_GW_Si_MnO_NiO_PRL2004} and $GW$@LDA+$U$ \cite{Kobayashi_GWLDAU_PRB2008} predictions of 2.42 eV and 2.27 eV, and 
\textit{iii}) an increase of the energy difference between the CBM at $\Gamma$ and the unoccupied $e_{g}$ bands, i.e. the bandgap.
 
The indirect bandgap improves from the PBE value of 0.98 eV to 2.31 eV; however, the experimental value of 3.6--4.1 eV (Table~\ref{tab:gaps}) is still underestimated. On the other hand, the magnetic moment is evaluated to be 4.79 $\mu_B$, which matches the experimental value {(Fender \textit{et al.} \cite{Fender_covalency_MnO_MnS_NiO_JCP1968}).}

Franchini {and coauthors} \cite{Franchini_manganese_oxides_PRB2007} have performed calculations with a larger value of $U_{3d} = 6.0$~eV ({without} $U$ on oxygen) yielding, however, a bandgap and magnetization (2.1 eV, 4.67$\mu_{B}$) smaller than our results. This evince{s} the importance of having the on-site Coulomb interaction not only on the TM but also on oxygen. Sakuma \textit{et al.}\cite{Sakuma_abinitio_dynamic_screening_TMO_PRB2013} have shown that O 2$p$ orbitals are considerably localized (as measured by the spread of their MLWFs) in TMOs; correspondingly, cRPA {\it ab-initio} calculations find the Coulomb repulsion in oxygen $\bar{U}_{2p} \gtrsim 4$ eV\cite{Shih_screened_interaction_TMO_PRB2012}, independent of the TM. Moreover, multideterminant correlated methods (complete active space self-consistent field, CASSCF) have recently shown that, contrary to conventional wisdom, the valence-band-edge in NiO is a localized O 2$p$ state\cite{alidoust2014revisiting_photoemission}.


\ACBN's MnO bandgap is closer to the predictions of hybrid functionals (sX-LDA 2.5 eV, HSE03 2.6 eV); nonetheless, the more accurate $G_0W_0$@HSE03 and self-consistent $GW$ methods yield results of 3.4 and 3.5 eV, much closer to the experimental range. This underperformance of the hybrid functionals suggests that correlation effects may be particularly predominant in the case of MnO and, therefore, beyond the Hartree-exchange correction of hybrid functionals.

\begin{figure}[htb]
\begin{center}
    \includegraphics[width=\columnwidth]{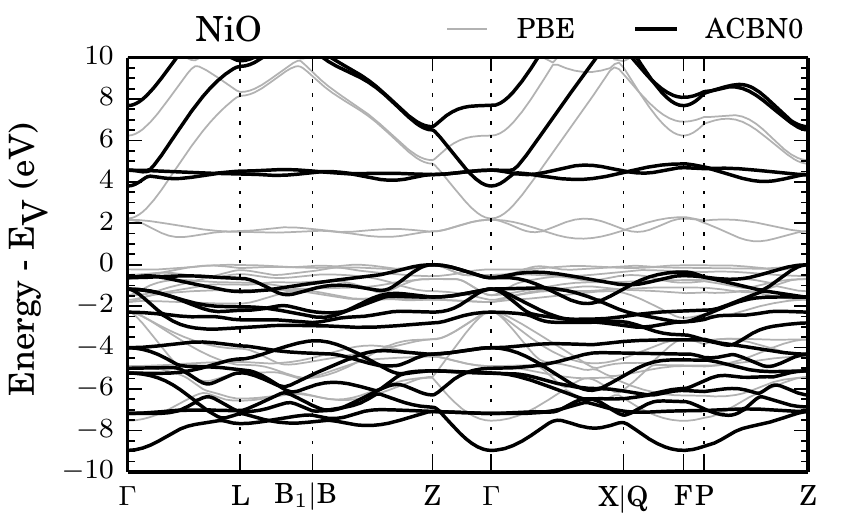}
    \vspace{-5mm}
    \caption{\small
      Band structure (spin up) of nickel oxide. All energies are relative to the valence band maximum E$_\textrm{V}$. 
      The effective values of $U_{3d}=7.63$ eV for nickel and $U_{2p}=3.0$ eV for oxygen are used in the \ACBN\, calculation.
} \label{fig:fig_nio} 
\end{center}
\end{figure}

For NiO, PBE incorrectly locates the parabolic 4$s$ band above the $3d$ ones, as seen in Fig.~\ref{fig:fig_nio}. With our values of $U$ 7.63 and 3.0 eV for Ni and O, the 4$s$ CBM is correctly positioned at the $\Gamma$ point \cite{Li_GW_NiO_PRB2005}, yielding an indirect bandgap (Z-$\Gamma$) of 3.8 eV and a direct gap of 4.29 eV. Considering that the 4$s$ CBM has low spectral weight, the direct gap can well account for the dominant first peak at 4.3 eV observed with bremsstrahlung isochromat spectroscopy (BIS) \cite{Sawatzky_gapNiO_PRL1984}.
Our value of Ni $U_{3d}=7.63$ eV is in the same range than the effective value reported by Anisimov {\it et al.} of 7.1 eV \cite{Anisimov_Mott_insulators_PRB1991}, obtained with the cLDA method, widely used for {\it ab-initio} calculation of $U$. Similar to the TiO$_2$ case, the bands around the bottom region of the NiO occupied manifold are strongly of O-$2p$ character and are noticeably downshifted by the use of O $U_{2p}$ with respect to the PBE bands. The manifold bandwidth increases by 1.5 eV to 9 eV as seen in Fig.~\ref{fig:fig_nio}. Most band structures reported in the literature do not find an increase of the bandwidth; nonetheless, the same value of 9 eV is obtained with the self-consistent $GW$ reported by Li {\it et al.} \cite{Li_GW_NiO_PRB2005} who argued that such broader bandwidth accounts well for the presence of strong satellite structures observed experimentally in that range of energy \cite{Fujimori_photoemission_nickel_compounds_PRB1984,Sawatzky_gapNiO_PRL1984}. Admittedly, these satellites are not captured in other approximations such as the model $GW$ of Massidda {\it et al.} \cite{Massidda_TMO_GW_PRB1997} As pointed by Gillen and Robertson \cite{Gillen_TMO_JPCM2013}, a bandwidth of 9 eV in NiO is in agreement with experimental measurements of 8--8.5 eV (X-ray emission spectroscopy XES \cite{Kurmaev_correlated_oxides_PRB2008}) and 8.5--9.5 eV (ultraviolet photoemission spectroscopy UPS \cite{Zimmermann_3d_TMP_columb_vs_covalency_JPCM1999}).
 
\begin{table}
\caption{\label{tab:magmom} Local magnetic moments (in $\mu_B$) for the antiferromagnetic states of MnO and NiO.}
\begin{ruledtabular}
\begin{tabular}{lcc}
   & MnO & NiO \\
\hline
PBE        & 4.58   & 1.49 \\
\rowcolor{Gray}
\ACBN\,    & 4.79   & 1.83 \\
HSE03      & 4.5 \cite{Rodl_quasiparticles_antiferro_oxides_PRB2009} & 1.5 \cite{Rodl_quasiparticles_antiferro_oxides_PRB2009}\\
$GW$@LDA & 4.6 \cite{Faleev_GW_Si_MnO_NiO_PRL2004}    & 1.9 \cite{Faleev_GW_Si_MnO_NiO_PRL2004} \\
Experiment & 4.58 \cite{Cheetham_magnetic_ordering_MnNiO_PRB1983}, 4.79 \cite{Fender_covalency_MnO_MnS_NiO_JCP1968} & 1.77 \cite{Fender_covalency_MnO_MnS_NiO_JCP1968}, 1.90 \cite{Cheetham_magnetic_ordering_MnNiO_PRB1983,Roth_magnetic_MnO_FeO_CoO_NiO_PR1958}
\end{tabular}
\end{ruledtabular}
\end{table}

\subsection{Wurtzite zinc oxide}

\begin{figure}[h]
    \includegraphics[width=\columnwidth]{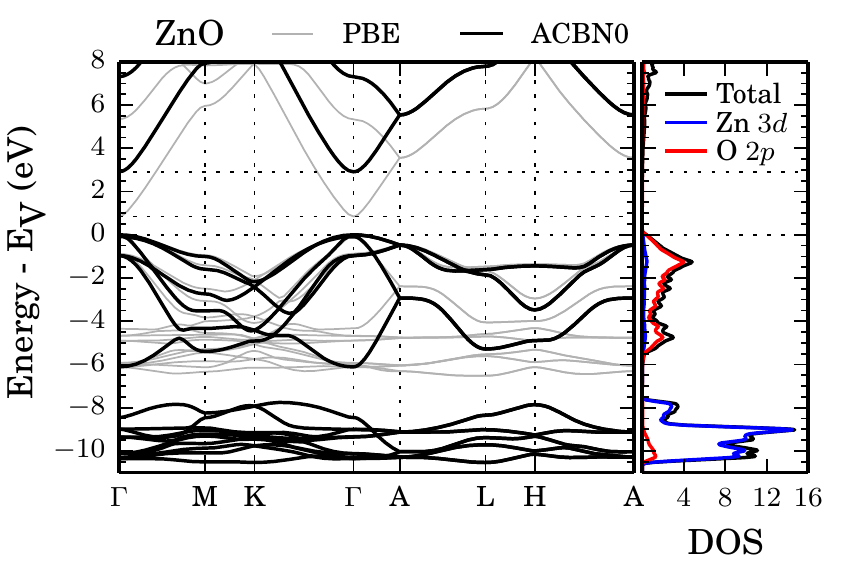}
    \caption{\label{fig:fig_zno}Comparison between the PBE (gray) and \ACBN\, (black) band structures for ZnO. The converged effective Coulomb interactions are Zn $U_{3d}=12.8$ eV and O $U_{2p}=5.29$ eV. The horizontal grid lines show the DFT (0.85 eV) and \ACBN\ (2.91 eV) bandgaps. The panel on the right shows the projected DOS for the \ACBN\, calculation}. 
\end{figure}

We use a hexagonal lattice (space group {\#}186) with relaxed lattice constants $a=b=3.1995$~\AA, $c=5.1330$~\AA, $\mu=0.3816$, 
 {taken from the AFLOWLIB database \cite{aflowlibPAPER} (Ref.~[\onlinecite{aflowAPI}], auid=aflow:b4819e0e63f994a8).}

In the strongly ionic ZnO, the bands can  be readily identified by their dominant orbital character. The bands in Fig.~\ref{fig:fig_zno} (black lines) from 0 to -6 eV are mostly of O-$2p$ character. The low-dispersion bands around -9 eV correspond to the Zn-$3d$ states. The conduction bands are predominantly of Zn-$4s$ character.

Within PBE, the $3d$ bands incorrectly overlap with the $2p$ manifold introducing spurious hybridizations, as shown in Fig.~\ref{fig:fig_zno} with gray lines, which in turn leads to a strong underestimation of the bandgap. The PBE gap is 0.85 eV while the experimental gap is 3.3 eV \cite{Dong_ZnO_PRB_2004}. Zinc oxide highlights the underlying failure of LDA or GGA in treating materials with localized
electrons and thus constitutes a case study for the application of the DFT+$U$ method. 

Our converged values are Zn $U_{3d} = 12.8$ eV and O $U_{2p}=5.29$ eV and yield a bandgap of 2.91 eV, which compares favorably to the experimental value. The bandwidth of the O-$2p$ manifold shown in Fig.~\ref{fig:fig_zno} is $\sim 6$ eV, in accordance to the angle-resolved photoemission spectroscopy (ARPES) value of $\sim 6.05$ eV \cite{Lim_arpes_calcu_ZnO_PRB2012}.

Although seemingly high, our parameters agree with values reported by Calzolari {\it et al.} \cite{Calzolari_ZnO_JACS2011} ($U_{3d}=12.0, U_{2p}=6.5$ eV) and Ma {\it et al.} \cite{Ma_ZnO_correlation_JPCC2013} ($U_{3d}=10, U_{2p}=7$ eV), both of which were found by fitting to reproduce the experimental bandgap and position of the $3d$ bands.

It is established that the $3d$ bands downshift monotonically with increasing values of $U_{3d}$. As the $3d$ bands downshift, the $p$--$d$ repulsion with the O-$2p$ bands is decreased, which in turns lowers the energy of the valence-band maximum (VBM) and, thus, monotonically increases the gap \cite{Jaffe_gap_chalcopyrite_PRB1984}. After the $3d$ bands have been fully disentangled from the $2p$ manifold, however, the $2p$ bands are well resolved and remain mostly insensitive to further increase of $U_{3d}$. {Consequently}, the bandgap becomes progressively independent of $U_{3d}$ and after the $3d$ bands are fully disentangled the application of on-site Coulomb interaction on oxygen becomes necessary to further reach the experimental bandgap \cite{Ma_ZnO_correlation_JPCC2013}. 

\begin{figure}[h]
    \includegraphics[width=\columnwidth]{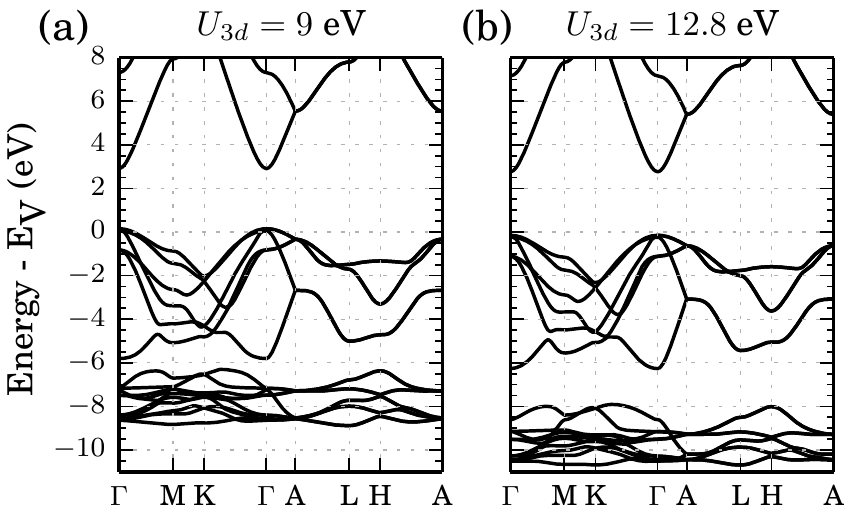}
    \caption{\label{fig:fig_zno_fixed_U} Effect of the zinc on-site Coulomb repulsion $U_{3d}$ on the occupied $3d$ bands. An \textit{ad hoc} value of $U_{3d}$=9.0 eV (a) is compared against the converged value of 12.8 eV (b). Both cases use the converged values of $U_{2d}$=5.29 eV for oxygen.} 
\end{figure}

For illustration, Fig.~\ref{fig:fig_zno_fixed_U} shows a comparison of the band structure with different values of $U_{3d}$ (12.8 and 9 eV), while the $U_{2d}$ is kept fixed at the converged value 5.29 eV. The unusual rigid shift of the $U_{3d}$ bands seen comparing Figs.~\ref{fig:fig_zno_fixed_U}(a) and (b) arises from a singularity particular to the case of the fully occupied $3d^{10}$ bands of Zn in ZnO that is rooted in the definition of the Hubbard correction to the energy functional: 
\begin{equation*} \label{eq:eu}
E_U=\frac{U}{2} \sum\limits_{I,\sigma} \sum\limits_{m}
\left[\lambda^{I\sigma}_m \left(1- \lambda^{I\sigma}_m \right) \right],
\end{equation*}
which is equivalent to Eq.~\ref{eq:E_U_dudarev} \cite{Cococcioni_DFTU_LinearResponse_PRB2005}, and the corresponding Hubbard potential is
\begin{equation*} \label{eq:Vu}
V_U=\frac{U}{2} \sum\limits_{I,\sigma} \sum\limits_{m}
\left(1- 2\lambda^{I\sigma}_m \right) |\phi^I_m\rangle \langle\phi^I_m|, 
\end{equation*}
where $0 \leq \lambda^{I\sigma}_m \leq 1$ is the occupation of the orbital $\phi_m$. For fully occupied orbitals such as Zn-$3d$ in ZnO (L\"{o}wdin charge 9.97$e$ out of 10$e$), i.e. $\lambda_m \approx 1$, the Hubbard energy reduces to $E_U \approx 0$, and the Hubbard potential becomes a rigid shift $V_U \approx -{U}/{2}$ applied to the localized orbitals. In principle, at this limit, the value of $U_{3d}$ does not change the energy of the material and {becomes irrelevant} in pinning the position of the $3d$ bands.

The experimental position of the center of the $3d$ bands is at -7.5 eV \cite{Powell_photoemission_ZnO_PRB1972,Lim_arpes_calcu_ZnO_PRB2012} measured with respect to the VBM (E$_{\textrm{V}}$). Other experimental values have also been reported -8.81 eV \cite{Ley_valence_band_density_III_V_II_VI_PRB1974}, -8.6 eV \cite{Vesely_core_states_IIB_VIA_PRB1971}, -7.8 eV \cite{Vogel_self_interaction_pseudopotentials_II_IV_PRB1996}. Our value of $U_{3d}$ underestimates the position of the $3d$ bands at $\sim$-9 eV. As discussed above, for fully disentangled and occupied $3d$ states, because a singularity of the DFT+$U$ energy functional, the energy of the system becomes almost independent of $U_{3d}$. 

Similarly, the $GW$ method and hybrid functionals, while correcting the bandgap and fully disentangling the $2p$ and $3p$ manifolds, consistently miss the position of the $3d$ bands by $\sim$ 1 eV \cite{Lim_arpes_calcu_ZnO_PRB2012}. Recently, Lim \textit{et al.} \cite{Lim_arpes_calcu_ZnO_PRB2012} proposed an assisted $GW+V_d$ approach in which the $3d$ bands are shifted by an \textit{ad hoc} on-site potential $V_d = 1.5$ eV in the $GW$ self-energy operator.

Analogously, the cLDA method fails in the case of Zn in ZnO. Because of the full occupancy of the $3d$ states, they become rather insensitive to small linear perturbations, yielding unreliable numerical values of $U$ \cite{Hu_LDAU_Co_ZnO_PRB2006,Cococcioni_reviewLDAU_2014}. Lee and Kim \cite{Lee_linear_response_Cu_Zn_ZnO_KOREA2012} have proposed an extension to cLDA method for systems with closed-shell localized electrons. They found $U_{3d}=5.4$ eV for Zn by applying a large perturbation potential and correcting for the excess potential needed to reach the onset of the electron-density response.

\section{Discussion} \label{sec:discussion}

Indeed, the \ACBN\, functional satisfies the rather ambitious criteria outlined by Pickett {\it et al.}\cite{Pickett1998}: 

{\bf I.} {\it \ACBN\, reduces to (LDA)PBE when (LDA)PBE is known to be good}.  
The reduction of $U\rightarrow 0$, arises with the introduction of the ``renormalized'' density-matrix $\bar{P}$ (instead of the regular density matrix $P$), which makes $U$ dependent on the degree of localization of the Bloch states.

A toy model with two basis functions $m$ and $m'$ reveals the scaling of Eq. 12 as
$\bar{U} \sim \frac{1}{4}N_{m}N_{m'} (mm|m'm')$ in contrast to $\bar{U} \sim (mm|m'm')$ when using the regular density matrix instead.
Delocalized Bloch states are assumed to be properly described at the LDA(PBE) level. The more delocalized a state, the lower the charge projected inside the atomic sphere ($N_m \approx N_{m'} \rightarrow 0$) and thereby $U$ vanishes quadratically. See for instance the case of Ti $U_{3d}$ in TiO$_2$, or silicon where \ACBN\, yields $U_{2p} \approx 0$ eV.

{\bf II.} {\it The energy is given as a functional of the density}. The value of $U$ in \ACBN\, depends only on the electron density, with a set of fixed PAOs $\{m\}$ chosen \textit{a priori} to define $U$. The \ACBN\, functional can be considered a zeroth-order pseudo-hybrid density functional in the sense that it includes an on-site form of the Hartree-Fock exchange. The results for the test systems studied here follow closely the more established and far computationally more expensive sX-LDA hybrid functional.
Moreover, the methodology presented in this work can be immediately generalized to evaluate the nonlocal exchange energy for solids by computing the full set of required two-electron integrals. Thus, one could have a hybrid-functional plane-wave DFT calculation that is as fast as LCAO hybrid-functionals while still benefiting from the robust parallel fast-Fourier-transform algorithms and systematic basis-set convergence of plane waves. 

{\bf III.} {\it The method specifies how to obtain the local orbital in question}.  
\ACBN\, directly parallels the original orbital-dependent DFT+$U$ functional of Anisimov 
that uses atomic orbitals $\{m\}$. 
Conceptually, the localized states $\varphi$ are linear combinations of $\{m\}$ 
with the expansion coefficients obtained self-consistently, thus, they reflect the chemical environment of the site; however, the expansion coefficients need not be explicitly known. Even though the information of the coefficients is conceptually included in the renormalized density matrix, they are not individually resolved.
Such LCAO expansion is more general and suitable for cases when
the localized bands are not readily disentangled from other bands, which happens when the orbitals $\varphi$ are not thoroughly well localized. 

{\bf IV.} {\it The definition of $\bar{U}$ and $\bar{J}$ is provided unambiguously}. See Eqs. \ref{eq:param_u} and \ref{eq:param_j}.

{\bf V.} 
{\it The method predicts antiferromagnetic insulators when appropriate.} 
{This is demonstrated by the results presented for TiO$_2$, MnO, NiO and ZnO.} 
The flexibility of ACBN0 is that it allows the calculation of  $\bar{U}$ and $\bar{J}$  for any atom in the system of interest, yielding for instance non-negligible values for the $2p$ lone-pair of oxygen in transition metal oxides or for the $p$ states of the anion in transition metal chalcogenides. Through the inclusion of these terms, ACBN0 corrects both the bandgap and the relative position of the different bands, in particular the ones deriving from the $d$ orbitals of transition metal atoms.  This characteristic of ACBN0 is crucial for improving the agreement with experimental results. Generally, the experimental bandgap of TMOs can not be reached if considering only the TM. Paudel and Lambrecht \cite{Paudel_corrected_O_ZnO_PRB2008} suggested the simultaneous use of $U$ on both the $3d$ and $4s$ orbitals of Zn to reach the experimental gap; however, a large value on Zn $U_{4s}=43.5$ eV was needed. Finally, our results predict the stability of the antiferromagnetic phases of both MnO and NiO. However, a more thorough discussion on the relative stability of different magnetic phases will be the subject of a forthcoming publication. \cite{PGopal_preparation_2014} 

\section{Conclusions} \label{sec:conclusions}

In conclusion, we have introduced \ACBN, a pseudo-hybrid density functional that incorporates the Hubbard correction of DFT$+U$ as a natural function of the electron density and chemical environment. The values of $\bar{U}$ and $\bar{J}$ are functionals of the electron density and provide a variational way of obtaining the proper description of insulators such as transition-metal oxides. Although a more extensive validation of this functional is needed, the first results of our tests show improved agreement to higher levels of theory (hybrid functionals, the \textit{GW} approximation) and to experimental measurements for the electronic properties of TMOs at a fraction of the computational cost. This is an essential requirement for the design efficient algorithms for electronic structure simulations of realistic material systems and massive high-throughput investigations \cite{nmatHT}.

\begin{acknowledgments}
We thank Drs. R. J. Mathar, N. J. Mosey, A. N. Andriotis, C. R\"{o}dl, A. Ferretti, A. Calzolari, P. Gopal, L. Liyanage, H. Shi, M. Fornari, 
G. Hart, {O. Levy, C. Toher, D. Irving}, and B. Himmetoglu for various technical discussions that have contributed to the results reported in this article.
This work was supported by ONR-MURI under contract N00014-13-1-0635 and the Duke University Center for Materials Genomics. 
S.C. acknowledges partial support by NIST (\#70NANB12H163) and DOE (DE-AC02-05CH11231, BES program under Grant \#EDCBEE).
We also acknowledge the Texas Advanced Computing Center (TACC) at the University of Texas Austin for providing HPC resources, 
 and the CRAY corporation for computational assistance.
 
\end{acknowledgments}

\appendix
\section{The PAO-3G minimal basis set}\label{sec:fitting}
The PAO basis functions $\phi_{lm}(\mathbf{r}) \equiv \frac{R_{l}(r)}{r} Y^{m\{c,s\}}_{l}(\theta,\varphi)$ are in fact obtained by solving the pseudopotential Kohn-Sham equation 
for a given atomic reference configuration, where $Y^{m\{c,s\}}_{l}$ are real-valued spherical harmonics \cite{Mathar_gaussianorbital_ijqm2002}.  
Given that the radial and angular part are separable, they can be directly fitted using linear combinations of spherical-harmonic Gaussian 
functions $\mathcal{G}_s(\mathbf{r},l,m,\zeta)=r^l e^{-{\zeta}r^2} Y_l^{m\{c,s\}}(\theta,\varphi)$. 
Then, $\phi(\mathbf{r})=\sum_{i=1}^{N_G} a_i \mathcal{G}_s(\mathbf{r},l,m,\zeta_i)$. 
The expansion coefficients $\{a\}$ and exponents $\{\zeta\}$ are found by fitting to $R_{l}(r_n)$, which is performed by using the non-linear least-square Levenberg–Marquardt 
algorithm to minimize the deviation 
\begin{equation} 
  \sum_{r_n} \left[ \sum_{i=1}^{N_G} a_i r^{l+1}_n e^{-{\zeta}_ir^2_n} - R_{l}(r_n) \right]^2\!\!.
\end{equation}
$R_{l}(r)$ is evaluated at a logarithmic radial mesh $\{r_n\}$ and provided in the atomic pseudopotential files taken from the 
PSlibrary 1.0.0 \footnote{\label{note} Pseudopotentials and PAOs used in this work are publicly available at \texttt{http://qe-forge.org/gf/project/pslibrary/}}. 
We only use norm-conserving pseudopotentials since they guarantee the charge conservation of $\phi$. 
The initial guess for the coefficients and exponents are taken from the STO-3G \cite{Hehre_self_consistent_molecular_orbit_JCP1969,Collins_self_consistent_molecular_orbit_JCP1976} basis-set, 
which associates three Gaussian functions per orbital ($N_G=3$), from the EMSL library \cite{Feller_role_databases_JCC1996}.

Traditionally, Cartesian Gaussian functions of the type $\mathcal{G}_c(\mathbf{r},l_x,l_y,l_z,\zeta) = x^{l_x} y^{l_y} z^{l_z} e^{-{\zeta} r^2}$ are held as the most efficient basis to compute the staggering number of two-electron integrals needed in quantum chemistry calculations. We follow the procedure by Mathar \cite{Mathar_gaussianorbital_ijqm2002,Mathar_ortho_ARXIV2009} to further convert each spherical-harmonic Gaussian into a linear combination of Cartesian Gaussians. Then, the Cartesian expansion of the PAOs is
\begin{equation} \label{eq:gauss_fit}
\phi_{lm}(\mathbf{r})=\frac{1}{4}\sqrt{\frac{(2l+1)!!}{\pi N_{l,m}}}f_{l,m}(x,y,z)\sum_{i=1}^{N_G} a_ie^{-\zeta_i r^2},
\end{equation}
where $f_{l,m}(x,y,z)$ is given in Table~\ref{tab:table1}.

\begin{figure}[htb]
  \begin{center}
    \includegraphics[width=0.99\columnwidth]{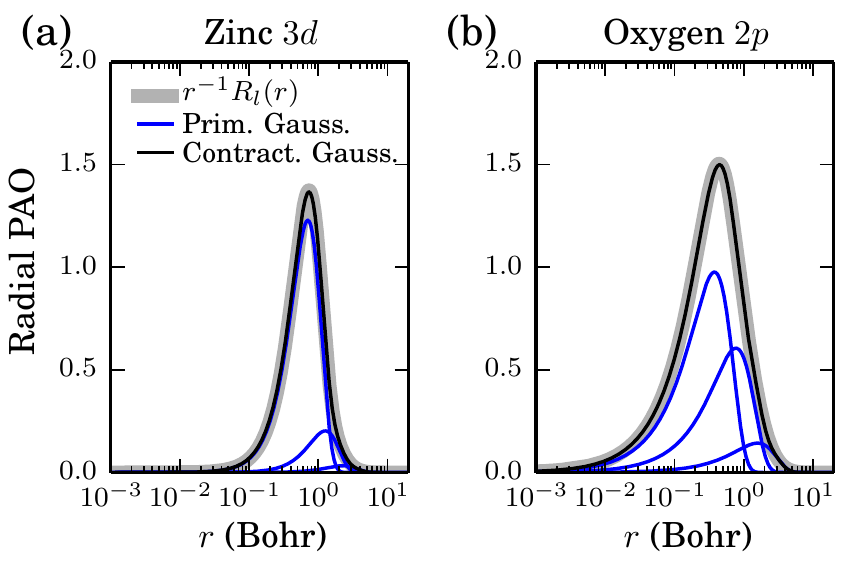}
    \vspace{-5mm}
    \caption{\small
      Fitting the radial component of the PAOs ($r^{-1} R_l(r)$, yellow line) as a linear combination of
      primitive Gaussians (black line). Each contracted Gaussian is the sum of the three primitive Gaussians 
      shown in blue. The Zn $d$ (a) and O $p$ PAOs (b) are taken from the PSlibrary 1.0.0.} 
    \label{fig:fig_5} 
  \end{center}
\end{figure}

An example of this fitting procedure is shown in Figure \ref{fig:fig_5} for Zn-$3d$ and O-$2p$ PAO. Having the PAOs expressed as linear combination of Gaussian-type orbitals in Eq.~\ref{eq:gauss_fit} is largely advantageous, since it allows computation of the ERIs in a straightforward and analytical way. Furthermore Gaussians allow filtering out ERIs with negligible energy contribution \cite{Izmaylov_shortrange_HF_JCP2006} further speeding up calculations, as implemented in the Heyd-Scuseria-Ernzerhof HSE03 \cite{Heyd2003} hybrid functional.  

\begin{table}[htb]
\caption{\label{tab:table1}
Cartesian expansion of $f_{l,m}$} 
\begin{ruledtabular}
\begin{tabular}{ C{3.0cm} C{3cm} c  }
$(l,m)$ & $N_{l,m}$ &  $f_{l,m}(x,y,z)$ \\ \hline
(0, 0) & 0.25& $1$   \\  \hline
(1,-1) & 0.25& $y $ \\
(1, 0) & 0.25& $z $    \\
(1, 1) & 0.25& $x $    \\ \hline
(2,-2) & 0.25& $xy $    \\
(2,-1) & 0.25& $yz $    \\
(2, 0) & 3   & $2z^2-x^2-y^2 $    \\
(2, 1) & 0.25& $xz $    \\
(2, 2) & 1   & $x^2-y^2 $    \\
\end{tabular}
\end{ruledtabular}
\end{table}

\section{Calculation of the real-space Hamiltonian and density matrices from the plane-wave electronic structure} \label{sec:hm_dm}

The solid is efficiently calculated using plane-wave DFT on a unit cell with periodic boundary conditions. 
The plane-wave basis allows a systematic convergence of the basis-set energy error, which is controlled by a single energy-cutoff parameter. 
Periodic-boundary conditions are implicit to the plane-wave basis, thus avoiding the presence of surface effects intrinsic to molecular cluster calculations. 
Moreover, plane waves allow the use robust and scalable Fourier-transform algorithms.
We follow the method described in Ref.~[\onlinecite{curtarolo:art86}] to project the $\mathbf{k}$-space electronic structure of the solid onto an atomic-orbital space by filtering 
out high-kinetic-energy plane waves. 
The resulting reciprocal-space Hamiltonian $H^{\sigma,\mathbf{k}}$ and overlap $S^{\mathbf{k}}$ matrices are then Fourier-transformed into real space resulting in:
\begin{equation} 
H^{\sigma,\mathbf{0R}}=\frac{1}{\sqrt{N_{\mathbf{k}}}} \sum\limits_{\mathbf{k}} e^{-i\mathbf{k}\cdot\mathbf{R}}{S^{\mathbf{k}}}^{\frac{1}{2}}H^{\sigma,\mathbf{k}}(\kappa,N){S^{\mathbf{k}}}^{\frac{1}{2}}, \label{eq:ft}
\end{equation}

\begin{equation} 
S^{\mathbf{0R}}=\frac{1}{\sqrt{N_{\mathbf{k}}}} \sum\limits_{\mathbf{k}} e^{-i\mathbf{k}\cdot\mathbf{R}}S^{\mathbf{k}}. \label {eq:ov} 
\end{equation}
The parameters $\kappa$ and $N$, defined in Ref.~\onlinecite{curtarolo:art86}, determine the shifting and filtering for the projection procedure. The overlap integral between a basis function $\phi_{\mu}$ located inside the primitive unit cell (lattice vector $\mathbf{0}$) and the periodic translation of $\phi_{\nu}$ to lattice vector $\mathbf{R}$ is the matrix element $S^{\mathbf{0R}}_{\mu \nu}=\langle {\mu}^{\mathbf{0}} | {\nu}^{\mathbf{R}}\rangle$.

The real-space density matrix is then computed as:
\begin{equation} \label {eq:dm} 
P^{\sigma,\mathbf{0R}}_{\mu\nu}=\frac{1}{\sqrt{N_{\mathbf{k}}}} \sum\limits_{\mathbf{k},i} e^{-i\mathbf{k}\cdot\mathbf{R}}N^{\mathbf{k\sigma}}_{\psi_i} c^{\mathbf{k\sigma}}_{\mu i}c^{\mathbf{k\sigma}}_{\nu i},
\end{equation}
where $N^{\mathbf{k\sigma}}_{\psi_i}=1$ for all occupied states $\psi^\mathbf{k\sigma}_i$. The expansion coefficients $c^{\mathbf{k\sigma}}_{\mu i}$ are the components of the generalized eigenvectors of $H^{\sigma,\mathbf{k}}$ and $S^{\mathbf{k}}$. 


\end{document}